\documentclass[preprint,1p,authoryear,number]{elsarticle}

\usepackage{amssymb, latexsym, amsthm, amsmath,lineno,epsfig,mathtools}

\newtheorem{thm}{Theorem}
\newtheorem{lem}[thm]{Lemma}
\newtheorem{cor}[thm]{Corollary}
\newdefinition{rmk}{Remark}
\newproof{pf}{Proof}

\newcommand\dbeta{\operatorname{beta}}

\newcommand\given{{\,|\,}}

\newcommand\eg{{\it e.g.,}}

\newcommand\ie{{\it i.e.,}}

\begin{document}
\begin{frontmatter}
\date{}

\title{Biallelic Mutation-Drift Diffusion in the Limit of Small Scaled
Mutation Rates}

\author{Claus Vogl\corref{cor1}}
\address[rvt]{Institute of Animal Breeding and Genetics, Veterin\"armedizinische
Universit\"at Wien, Veterin\"arplatz 1, A-1210 Vienna, Austria\corref{cor1}}
\cortext[cor1]{Corresponding author}
\ead{claus.vogl@vetmeduni.ac.at}

\begin{abstract}
  The evolution of the allelic proportion $x$ of a biallelic locus
  subject to the forces of mutation and drift is investigated in a
  diffusion model, assuming small scaled mutation rates. The overall
  scaled mutation rate is parametrized with $\theta=(\mu_1+\mu_0)N$
  and the ratio of mutation rates with
  $\alpha=\mu_1/(\mu_1+\mu_0)=1-\beta$. The equilibrium density of
  this process is beta with parameters $\alpha\theta$ and
  $\beta\theta$. Away from equilibrium, the transition density can be
  expanded into a series of modified Jacobi polynomials. If the
  parameters $\alpha$ or $\theta$ change, this eigenexpansion also has
  to change, such that modeling, \eg\ growing or shrinking populations
  is cumbersome.---If the scaled mutation rates are small, i.e.,
  $\theta \ll 1$, it may be assumed that polymorphism derives from
  mutations at the boundaries. A model, where the interior dynamics
  conform to the pure drift diffusion model and the mutations are
  entering from the boundaries is derived. In equilibrium, the density
  of the proportion of polymorphic alleles, \ie\ $x$ within the
  polymorphic region $[1/N,1-1/N]$, is inversely related to the
  distance from the origin at the boundaries and symmetric
  $\alpha\beta\theta(\tfrac1x+\tfrac1{1-x})=\tfrac{\alpha\beta\theta}{x(1-x)}$,
  while the mutation bias $\alpha$ influences the proportion of
  monomorphic alleles at 0 and 1. Analogous to the expansion with
  modified Jacobi polynomials, a series expansion of the transition
  density is derived, which is connected to Kimura's well known
  solution of the pure drift model using Gegenbauer polynomials. Two
  temporal and two spatial regions are separated. The eigenvectors
  representing the spatial component within the polymorphic region
  depend neither on the on the scaled mutation rate $\theta$ nor on
  the mutation bias $\alpha$. Therefore parameter changes, e.g.,
  growing or shrinking populations or changes in the mutation bias,
  can be modeled relatively easily, without the change of the
  eigenfunctions necessary for the series expansion with Jacobi
  polynomials. With time, the series converges to the equilibrium
  solution.
\end{abstract}


\end{frontmatter}


\section{Introduction}

In this manuscript, it is assumed that the proportion $x$ in the
population of the first allelic type of a biallelic locus is evolving
independently according to a biallelic mutation drift model
\citep[e.g.][]{Wrig31,Ewen04,Grif10,Song12}. In the diffusion limit,
the biallelic mutation drift model is usually parametrized with the
two parameters $\theta_1=\mu_1N$ and $\theta_0=\mu_0N$, where $\mu_1$
and $\mu_0$ are the mutation rates towards alleles one and zero,
respectively, and $N$ is the haploid effective population number or
size. For small scaled mutation rates, polymorphism probably derives
from a mutation at the boundaries and the analysis simplifies. The
following reparametrization is then convenient:
$\alpha=\mu_1/(\mu_1+\mu_0)=1-\beta$ and
$\theta=\theta_1+\theta_0$. According to simulations partially
published in \citet{Vogl12} (see their Fig.~1, for a polymorphic
sample), this simplification holds for $2 \theta_0
\theta_1/(\theta_0+\theta_0)<0.02$ or $\alpha\beta\theta<0.01$. Note
that this assumption of small scaled mutation rates was already
discussed by \citet{Wrig31} and underlies much of population genetic
theory, e.g., the derivation of the Ewens-Watterson estimator of
$\theta$ \citep{Ewen04,Watt75} or Poisson Random Field approaches
\citep[\eg][]{Sawy92,RoyC10}.

Assuming a single segregating mutation and thus, implicitly, small
scaled mutation rates, alleles can be polarized into ancestral and
derived with information from related species or populations (outgroup
information). In the absence of selection and for constant $N$, the
density of the proportion of mutant polymorphic alleles $x$ converges
to be inversely related to the distance from the ancestral state, \ie\
proportional to $1/x$ or $1/(1-x)$, depending on the ancestral state
\citep{Wrig31}. In equilibrium, equal amounts of mutant alleles
originate from the two boundaries \citep{Wrig31}, such that the
density of the proportion of polymorphic alleles becomes proportional
to $1/x+1/(1-x)=1/(x(1-x))$.

While the small scaled mutation rate assumption has been very
important in population genetic theory, particularly with data
analysis, only few attempts have been made to link the model with
general mutation rates to one with small scaled mutation
rates. \citet{Gute09} present a model for the analysis of site
frequency spectra that considers two parameter regions. Within the
polymorphic region, i.e., between $1/N$ and $1-1/N$ allelic
proportions evolve according to a selection, migration, and drift
model. Mutations are considered as follows \citep{Gute09}: ``Because
the diffusion equation [incorporating selection, migration, and drift]
is linear, we can solve simultaneously for the evolution of all
polymorphism by continually injecting $\phi$ density at low frequency
in each population (at a rate proportional to the total mutation flux
$\theta$), corresponding to novel mutations.'' Mutations are assumed
to only arise at the boundaries, presumably in equal proportions. The
authors do not justify this assumption any further.---This model of
mutations from only the boundaries is essentially the one considered
in this article. In contrast to \citet{Gute09}, who use a grid based
numerical approach for solving the diffusion equation, herein, changes
in the mutation bias are allowed and orthogonal polynomials are
used. The latter are exact, if assumptions are met, and offer a
connection to other theoretical work.

Independently from \citet{Gute09}, \citet{Vogl12} analyzed a Moran
model of mutation, selection, and drift and motivated a simpler model
with mutations only entering from the boundaries. This assumption was
justified by the observation that in equilibrium each particle spends
only a proportion of time in the polymorphic region of approximately
$2\alpha\beta\theta\log(N)$, which is small unless $N$ is very
large. With the diffusion model, however, the limit of the population
size to infinity $N\to\infty$ is considered. This makes the above
argument obsolete and necessitates a new justification, which will be
provided herein.

\paragraph{Outlook} 
First, the general biallelic mutation and drift Moran model and the
corresponding forward diffusion model will be reviewed, which can be
solved using a series expansion of (modified) Jacobi polynomials
\citep{Grif10,Song12,Vogl14}. Then the assumption of small scaled
mutation rates will be introduced, the modified Moran model with
mutations only from the boundaries will be reviewed, and the
corresponding diffusion model will be derived. A dynamical system
using orthogonal Gegenbauer polynomials will be motivated. This system
converges to an equilibrium solution with time. This equilibrium
density will be compared to the general equilibrium solution. Finally,
an example involving a change in the mutation bias will be shown.

\section{The General Mutation-Drift Model}

\subsection{Moran and Diffusion Models}

Assume a population of $N$ haploid individuals; each may assume the
state of zero or one, corresponding to the two arbitratrily labeled
alleles. With the decoupled Moran model \citep{Baak08,Ethe09,Vogl12},
either i) ({\bf mutation}) at a rate of $\mu=\mu_0+\mu_1$, a random
individual $i$ is picked to mutate to type one with probability
$\mu_1/\mu$ or to type zero with probability $\mu_0/\mu$; or ii) ({\bf
  genetic drift}) at a rate of one, a random individual $i$ is
replaced by another random individual $j$.  Thus, the rate of change
of the allelic proportion $x$ per unit time of the mean is caused by
mutation
\begin{equation}
  \operatorname{M}_{\delta x}=\frac1{N^2}\theta (\alpha-x) N\,,
\end{equation}
and that of the variance by genetic drift
\begin{equation}
  \operatorname{V}_{\delta x}=\frac2{N^2} x(1-x) N^2 \,.
\end{equation}
Scaling space with $1/N$ and time with $1/N^2$ and taking the
appropriate limits, the Kolmogorov forward (or Fokker-Planck)
generator of the process becomes
\begin{equation}\label{eq:Kol_for}
  {\cal L}_f=\left(\frac{\partial^2}{\partial x^2}x(1-x)\right)-\left(\frac{\partial}{\partial x}\theta(\alpha-x)\right)\,.
\end{equation}
The forward diffusion equation
\begin{equation}\label{eq:Kol_for_diff}
 \frac{\partial}{\partial  t} \phi(x,t)= {\cal L}_f \phi(x,t)
\end{equation} 
then describes the evolution of the probability of the allelic proportion
$x$ forward in time $t$. This is the same temporal direction as the
transitions in the Wright-Fisher and Moran models.

\subsection{Modified Jacobi Polynomials}

For the following, we will briefly recapitulate the theory of
orthogonal polynomials; a more detailed review can be found in
\citet{Vogl14}. 

On the interval $[0,1]$ we are looking for solutions of the Kolmogorov
forward equation Substituting the function
$\phi(x,t)=\sum_{i=0}^\infty e^{-\lambda_i t}w(x)\,f_i(x)$ into the
Kolmogorov forward equation, results in
\begin{equation}\label{eq:forw}
     -\lambda_i w(x) f_i(x)=\left(\frac{d^2}{d x^2}x(1-x) w(x) f_i(x)\right) 
    +\left(\frac{d}{dx}\theta(\alpha-x) w(x)f_i(x)\right)\,,
\end{equation}
where $i$ indexes the eigenvectors and $w(x)$ is the weight function
\begin{equation}\label{eq:weight}
  w^{(\theta,\alpha)}(x)=x^{\alpha\theta-1}(1-x)^{\beta\theta-1}\,.
\end{equation}
It can be shown that all eigenvectors are real and can be ordered such
that $\lambda_0<\lambda_1<\lambda_2<\cdots<\lambda_i<\cdots \to
\infty$. Corresponding to each eigenvalue $\lambda_i$ is a unique (up
to a normalization constant) eigenfunction $f_i(x)$, which has exactly
$i$ zeros in the interval.

This solution of the Kolmogorov forward equation (\ref{eq:forw}) can
be algebraically transformed to a solution of the corresponding
Kolmogorov backward equation
\begin{equation}\label{eq:backward}
     -\lambda_i f_i(x)=\left(x(1-x)\frac{d^2}{d x^2}f_i(x)\right) 
    +\left(\theta(\alpha-x)\frac{d}{dx} f_i(x)\right)\,.
\end{equation}
This backward equation (\ref{eq:backward}) is closely related to the
differential function fulfilled by the classical Jacobi polynomials
\citep{Abra70}. Define the modified Jacobi polynomials \citep{Song12}
\begin{equation}\label{eq:modified_jacobi}
  R_i^{(\theta,\alpha)}(x)=P_i^{(\beta\theta-1,\alpha\theta-1)}(2x-1)\,,
\end{equation}
where $P_i^{(a,b)}(z)$ are the classical Jacobi
polynomials \citep{Abra70}. It can be shown that these modified Jacobi
polynomials fulfil the backward equation (\ref{eq:backward}) with the
corresponding eigenvalues
\begin{equation}
  \lambda_i=i(i+\theta-1)\,.
\end{equation}

With the weight function $w^{(\theta,\alpha)}(x)$, the modified Jacobi
polynomials are orthogonal:
\begin{equation}\label{eq:ortho_rel}
  \int_0^1  R_i^{(\theta,\alpha)}(x) R_j^{(\theta,\alpha)}(x)\,w^{(\theta,\alpha)}(x)\, dx =  \Delta_i^{(\alpha,\theta)}\delta_{i,j}\,,
\end{equation}
where $\delta_{i,j}$ denotes the Kronecker delta, i.e., $\delta_{i,j}$
is zero for $i\neq j$ and one for $i=j$. The proportionality constant
$\Delta_i^{(\alpha,\theta)}$ is finite
\begin{equation}\label{eq:prop_const}
  \Delta_i^{(\alpha,\theta)}=\frac{\Gamma(i+\alpha\theta)\Gamma(i+\beta\theta)}{(2i+\theta-1)\Gamma(i+\theta-1)\Gamma(i+1)}\,.
\end{equation}
The set of $R_i^{(\theta,\alpha)}(x)$ forms a basis of the Hilbert
space $L^2([0,1])$ with the weight function $w^{(\theta,\alpha)} (x)$
\citep{Song12}.

For $i\geq 1$, the $R_i^{(\theta,\alpha)}(x)$ satisfy the recurrence
relation
\begin{equation}\label{eq:rec_rel}
  \begin{split}
   &R_{i+1}^{(\theta,\alpha)}(x)\frac{(i+1)(i-1+\theta)}{(2i+\theta)(2i-1+\theta)}=\\
    &\qquad R_i^{(\theta,\alpha)}(x)\left(x-\tfrac12+\frac{\theta^2(\beta^2-\alpha^2)-2
        \theta (\beta-\alpha)}{2(2i+\theta)(2i-2+\theta)}\right)\\
    &\qquad -R_{i-1}^{(\theta,\alpha)}(x)\frac{(i-1+\alpha \theta)(i-1+\beta \theta)}{(2i-1+\theta)(2i-2+\theta)}\,,\\ 
  \end{split}
\end{equation}
while $R_0^{(\theta,\alpha)}(x)=1$ and
$R_1^{(\theta,\alpha)}(x)=\theta(x-\alpha)$  \citep{Song12}. 

If $\theta > 0$, the forward equation has a stationary beta density
proportional to the weight function:
\begin{equation}\label{eq:equil_beta}
  \begin{split}
    f(x\given \theta,\alpha,\beta, t\to\infty) 
    &= \frac{\Gamma(\theta)}{\Gamma(\alpha\theta)\Gamma(\beta\theta)}
    \, w^{(\theta,\alpha)}(x)\, R_0^{(\theta,\alpha)}(x)
    =\frac{\Gamma(\theta)}{\Gamma(\alpha\theta)\Gamma(\beta\theta)}\,x^{\alpha\theta-1}(1-x)^{\beta\theta-1}\\
    &=\dbeta(x\given \alpha\theta,\beta\theta)\,.
  \end{split}
\end{equation}

The evolution of $x$ forward in time is given by the expansion:
\begin{equation}\label{eq:expansion}
  f(x\given \theta,\alpha,\beta, t)= w^{(\theta,\alpha)}(x)
  \left(c_0+\sum_{i=1}^{\infty} e^{-i(i+\theta-1)\, t}\,c_i\,R_i^{(\theta,\alpha)}(x)\right)\,.
\end{equation}
In practice, the expansion needs to be terminated at a finite $n$.
The constants $c_i$ are determined such that the initial conditions
are met, i.e., an initial probability density $f(x)$, defined within
the interval, is represented by the series expansion
\begin{equation}
  f(x)= w^{(\theta,\alpha)}(x) \left(c_0+\sum_{i=1}^{n} c_iR_i^{(\theta,\alpha)}(x)\right)\,.
\end{equation}
By minimizing the weighted least squares error function
\begin{equation}
  E(c_0,\dots,c_n)=\int_0^1 w(x)^{-1} \left(f(x) - \sum_{i=0}^n c_i w(x)^{(\theta,\alpha)} R_i^{(\theta,\alpha)}(x)\right)^2 dx\,.
\end{equation}
the coefficients are determined to be
\begin{equation}\label{eq:inner_forw}
  c_i=\frac1{\Delta_i}\int_0^1 R_i^{(\theta,\alpha)}(x) f(x)\,dx\,.
\end{equation}
Often an initial density corresponding to a Dirac delta
function at a point $p$ in $[1/N,1-1/N]$, $f(x)=\delta(x-p)$, is considered
\citep[e.g.,][]{Kimu55}. Then the expansion becomes
\begin{equation}
  f(x\given \theta,\alpha,p,t)= w^{(\theta,\alpha)}(x)
  \left(c_0+\sum_{i=1}^{n} e^{-i(i+\theta-1)\, t}\,\,R_i^{(\theta,\alpha)}(x) \frac{R_i^{(\theta,\alpha)}(p)}{\Delta_i^{(\theta,\alpha)}}\right)\,.
\end{equation}
This corresponds to formula (4.68) in \citet{Ewen04}, where
$n\to\infty$ and the eigenfunctions are assumed to be normed, such
that division by the proportionality constant
$\Delta_i^{(\alpha,\theta)}$ is unnecessary.

From the orthogonality relation (\ref{eq:ortho_rel}) and
$R_0^{(\theta,\alpha)}(x)=1$, it can be deduced for all $i\geq 1$ and
thus also for all times
\begin{equation}
  \begin{split}
    0&=\int_0^1 R_i^{(\theta,\alpha)}(x)\,R_0^{(\theta,\alpha)}(x)\,w(x)\,dx
    =\int_0^1 R_i^{(\theta,\alpha)}(x)\, w^{(\theta,\alpha)}(x)\,dx\,.
  \end{split}
\end{equation}
Therefore the probability mass over the whole interval $[0,1]$ comes
only from the equilibrium term, i.e., the beta density
(\ref{eq:equil_beta}); all other terms $R_i^{(\theta,\alpha)}(x)\,
w^{(\theta,\alpha)}(x)$ with $i\geq 1$ shift this mass within the
interval. 

\paragraph{Expression of the modified Jacobi polynomials as linear
  combinations of Beta densities} Note that a polynomial times a
beta results in a weighted sum of beta densities. This can be made
even more explicit by using the following representation of the
modified Jacobi polynomials \citep[compare][22.3.1]{Abra70}
\begin{equation}\label{eq:explicit_jacobi}
  R_i^{(\theta,\alpha)}(x)=\sum_{m=0}^i \frac{(-1)^{i-m}\Gamma(i+\alpha\theta) \Gamma(i+\beta\theta) }{\Gamma(i-m+1)\Gamma(m+\alpha\theta) \Gamma(m+1)\Gamma(i-m+\beta\theta) }\,x^m(1-x)^{i-m}
\end{equation}
to obtain
\begin{equation}\label{eq:Jacobi_as_betas}
  \begin{split}
    w^{(\theta,\alpha)}(x)\,R_i^{(\theta,\alpha)}(x) &=
    \sum_{m=0}^i \frac{(-1)^{i-m}\Gamma(i+\alpha\theta) \Gamma(i+\beta\theta) }{\Gamma(i-m+1)\Gamma(m+\alpha\theta)\Gamma(m+1)\Gamma(i-m+\beta\theta) }\\
    &\qquad\cdot x^{m+\alpha\theta-1}(1-x)^{i-m+\beta\theta-1}\\
    &=\sum_{m=0}^i \frac{(-1)^{i-m}\Gamma(i+\alpha\theta)
      \Gamma(i+\beta\theta)
    }{\Gamma(i-m+1)\Gamma(m+1)\Gamma(i+\theta)}\,\dbeta(x\given m+\alpha\theta,i-m+\beta\theta)\,.
  \end{split}
\end{equation}

\subsubsection{Data: Likelihood, Joint and Posterior Densities,
  and the Marginal Distribution with Modified Jacobi Polynomials}

While often a Dirac delta starting density was considered
\citep[eg.,][]{Kimu55,Ewen04}, we will usually have a sample of size
$M$ with $y$ alleles of type one. Given the allelic proportion $x$ the
distribution of alleles is naturally modeled as a binomial
\begin{equation}\label{eq:binomial}
  \Pr(y\given x,M)=\binom{M}{y}\,x^y(1-x)^{M-y}\,.
\end{equation}
The joint density of $y$ and $x$ after multiplication with the
equilibrium beta density (\ref{eq:equil_beta}) is
\begin{equation}\label{eq:joint}
  \Pr(y,x\given \alpha,\theta,M)=
  \binom{M}{y}\,\frac {\Gamma(\theta)}{\Gamma(\alpha\theta)\Gamma(\beta\theta)}\,
    x^{y+\alpha\theta-1}(1-x)^{M+\beta\theta-y-1}\,.
\end{equation}
Integrating out $x$ results in the beta-binomial compound distribution
\begin{equation}\label{eq:beta_binomial}
  \begin{split}
  \Pr(y\given \alpha,\theta,M)&=
  \binom{M}{y}\, \frac{\Gamma(\theta)}{\Gamma(\alpha\theta)\Gamma(\beta\theta)}
    \int_0^1 x^{y+\alpha\theta-1}(1-x)^{M-y+\beta-1}\, dx \\
    &=\binom{M}{y}\,
    \frac{\Gamma(\theta)}{\Gamma(\alpha\theta)\Gamma(\beta\theta)}\,
    \frac{\Gamma(y+\alpha\theta)\Gamma(M-y+\beta\theta)}{\Gamma(M+\theta)}\,.
  \end{split}
\end{equation}
The posterior of $x$ (i.e., the conditional probability density of $x$
after observing the data $y$ given $M$) is a beta density
\begin{equation}\label{eq:post}
  \begin{split}
    \Pr(x\given \alpha,\theta,y,M)&=
    \frac {\Gamma(M+\theta)}{\Gamma(y+\alpha\theta)\Gamma(M-y+\beta\theta)}\,
    x^{y+\alpha\theta-1}(1-x)^{M-y+\beta\theta-1}\\
    &=\dbeta(x\given y+\alpha\theta,M-y+\beta\theta)\,.
  \end{split}
\end{equation}

\subsubsection{Example: A Change in the Mutation Bias with Modified
Jacobi Polynomials} 

As an example, assume that the population had been in equilibrium with
parameters $\alpha_a$ and $\theta$, to switch to a new mutation bias
$\alpha_c$ at time $t_c$, while $\theta$ has remained constant
throughout. Then the expansion until time $t_c$ contains only the
equilibrium beta density. The change of the mutation bias
necessitates a change in the eigenvectors from
$w^{(\theta,\alpha_a)}\, R_i^{(\theta_a,\alpha)}$ to
$w^{(\theta,\alpha_c)}\, R_i^{(\theta,\alpha_c)}$. The coefficients
for the new eigensystem are (compare formula~\ref{eq:inner_forw})
\begin{equation}\label{eq:change_jacobi}
  c_i=\frac1{\Delta_i}\int_0^1 R_i^{(\theta,\alpha_c)}(x) w^{(\theta_a,\alpha)}\,
R_0^{(\theta,\alpha_a)}\,dx\,.
\end{equation}
The evolution of the proportion $f(x)$ between $t_c$ and the present
time is given by the series expansion (\ref{eq:expansion}) with the
$c_i$ from equation (\ref{eq:change_jacobi}).

While one such change may not be too cumbersome to implement in a
computer program, approximating, e.g., exponentially growing or
shrinking populations by many piecewise linear changes can be if
equilibrium has not been reached, since then
for each change a sum over all terms in the expansion is needed and
equation (\ref{eq:change_jacobi}) needs to be modified to
\begin{equation}
  c_i=\frac1{\Delta_i}\int_0^1 R_i^{(\theta,\alpha_c)}(x) w^{(\theta_a,\alpha)}\,
\sum_i R_i^{(\theta,\alpha_a)}\,dx\,.
\end{equation}
A substantial improvement can be the use of the assumption of
mutations only from the boundaries, where such a change of the
eigensystem is not necessary. This will be investigated in the next
section.

\section{Mutation-Drift With Small Scaled Mutation Rates}

\subsection{Pure Drift Diffusion}

In this subsection, the pure drift diffusion model is reviewed, as it
is the basis for the boundary mutation-drift model. In the interior,
i.e., inside the polymorphic region between $[1/N,1-1/N]$, the
dynamics of the allelic proportion $x$ are influenced only by drift,
such that the forward generator simplifies to
\begin{equation}
  {\cal L}_f=\frac{\partial^2}{\partial x^2}x(1-x)\,,
\end{equation}
and the corresponding Kolmogorov forward equation to
\begin{equation}\label{eq:forward_nomut}
  \frac{\partial}{\partial t} \phi(x,t)= {\cal L}_f  \phi(x,t)
  =\frac{\partial^2}{\partial x^2}x(1-x)  \phi(t,x)\,.
\end{equation}
Note that with the general mutation drift Kolmogorov forward equation
(\ref{eq:Kol_for_diff}) the boundaries are regular, i.e., accessible
and non-absorbing, whereas with the pure drift model the boundaries
are usually considered exit boundaries, i.e., accessible and absorbing
\citep{Ewen04}.

The dynamics of the polymorphic region have been analyzed by
\citet{Kimu55} and \citet{Song12} using Gegenbauer polynomials
\citep[e.g.,][]{Kimu55,Ewen04,Song12}.  \citet{Tran14b} suggested to
augment the eigenvectors by boundary terms, which results in a
``global'' solution that, in addition to the polymorphic region within
$[1/N,(N-1)/N]$ includes the boundaries zero and one. We will follow
this strategy, while maintaining the connection to the modified Jacobi
polynomials \citet{Song12} as defined in (\ref{eq:modified_jacobi}). 

For $i\geq2$, define the following set of orthogonal polynomials with
boundary terms:
\begin{equation}\label{eq:fast_eigenvectors}
  H_i(x)= \frac{(-1)^{i}\,\delta(x)+\delta(x-1)}{i}+U_i(x)
  \,,
\end{equation}
with 
\begin{equation}
  U_{i+2}(x)=x^{-1}(1-x)^{-1}G_{i}(x)=-\frac{2}{i+2}C_i^{(3/2)}(2x-1)=R^{(\theta=2,\alpha=1/2)}\,,
\end{equation}
where the $G_i(x)$ are the modified Gegenbauer polynomials
\citep{Song12}, the $R^{(\theta=2,\alpha=1/2)}$ are defined in
(\ref{eq:modified_jacobi}) \citep{Song12}, and the
$C_i^{(\alpha)}(z)$ correspond to the classical ultraspherical or
Gegenbauer polynomials with $\alpha=3/2$ \citep[][chap.22]{Abra70}
used by \citet{Kimu55}. 

Note that, for $i\geq2$, the boundary terms of $H_i(x)$, i.e., the Dirac
delta function, at zero and one are
\begin{equation}\label{eq:bound_vals}
  \begin{cases}
    \int_0^1 x U_i(x)\,dx=1/i\\
    \int_0^1 (1-x) U_i(x)\,dx=(-1)^i/i\,.
  \end{cases}
\end{equation} 

\paragraph{Expression of the modified Gegenbauer polynomials as linear
  combinations of Beta densities} 
The modified Gegenbauer polynomials can be represented explicitly as
polynomials and also as linear combinations of beta densities, as with
the modified Jacobi polynomials (eq.~\ref{eq:Jacobi_as_betas}):
\begin{equation}\label{eq:Gegenbauer_as_betas}
  \begin{split}
    U_{i+2}(x) &=\sum_{m=0}^i (-1)^{i-m+1}\frac{(i+1)!}{m!(i-m+1)!)}\frac{(i+1)!}{(m+1)!((i-m)!}x^{m}(1-x)^{i-m}\\
    &=
    \sum_{m=0}^i
    (-1)^{i-m+1}\frac{(i+1)!}{(m+1)!(i-m+1)!}\,\dbeta(x\given m+1,i-m+1)\,.
    \end{split}
\end{equation}
In this case, the beta densities have integer parameters greater than
one, i.e., are polynomials.

\begin{lem}
  The set of eigenvectors $H_i(x)$, for $i\geq 2$, can be derived from
  the modified Jacobi polynomials in equation
  (\ref{eq:modified_jacobi}) \citep{Song12} multiplied by the weight
  function, $w^{(\theta,\alpha)}(x)R_i^{(\theta,\alpha)}(x)$, if i)
  only terms in a Taylor expansion in $\theta$ up to zeroth order are
  kept in the polymorphic region $]0,1[$, while ii) terms that, for
  $\theta\to0$, vanish in the interior and converge to point masses at
  the boundaries are set to those values there; compactly,
  \begin{equation}\label{eq:power_jac_gegen}
    w^{(\theta,\alpha)}(x)R_{i}^{(\theta,\alpha)}(x)=H_i(x) +O(\theta)\,.
  \end{equation}
\end{lem}

\begin{pf}
  For $i\geq 1$,
  \begin{equation}
    \begin{split}
      w^{(\theta,\alpha)}(x)R_{i}^{(\theta,\alpha)}(x)
      &=    \sum_{m=0}^i \frac{(-1)^{i-m}\Gamma(i+\alpha\theta) \Gamma(i+\beta\theta) }{\Gamma(m+1) \Gamma(i-m+1)\Gamma(m+\alpha\theta) \Gamma(i-m+\beta\theta)}\\
      &\qquad\cdot x^{m+\alpha\theta-1}(1-x)^{i-m+\beta\theta-1}\\
      &=\sum_{m=1}^{i-1} \frac{(-1)^{i-m}\Gamma(i) \Gamma(i)}{\Gamma(m+1)\Gamma(i-m) \Gamma(i-m+1)\Gamma(m)}\\
      &\qquad\cdot x^{m-1}(1-x)^{i-m-1}
      +(-1)^{i}\delta(x)/i+\delta(x-1)/i +O(\theta)\\
      &=\sum_{m=0}^{i-2} \frac{(-1)^{i-m-1}\Gamma(i) \Gamma(i)}{\Gamma(m+2)\Gamma(i-m-1) \Gamma(i-m)\Gamma(m+1)}\\
      &\qquad\cdot x^{m}(1-x)^{i-m-2}+(-1)^i\delta(x)/i+\delta(x-1)/i +O(\theta)\\
      &=\sum_{m=0}^{j} \frac{(-1)^{j-m+1}\Gamma(j+2) \Gamma(j+2)}{\Gamma(m+2)\Gamma(j-m+1) \Gamma(j-m+2)\Gamma(m+1)}\\
      &\qquad\cdot x^{m}(1-x)^{j-m} +(-1)^{i}\delta(x)/i+\delta(x-1)/i \\
      &=\sum_{m=0}^{j} \frac{(-1)^{j-m+1}(j+1)! (j+1)!}{(m+1)!(j-m)!(j-m+1)!m!}\,x^{m}(1-x)^{j-m}\\
      &\qquad+(-1)^{i}\delta(x)/i+\delta(x-1)/i\\
      &=U_{i}(x) +(-1)^{i}\delta(x)/i+\delta(x-1)/i +O(\theta)\\
      &=H_i(x) +O(\theta)\,,
    \end{split}
  \end{equation}
  where $j=i-2$.
\end{pf}

\begin{rmk}
  The $H_i(x)$ are obviously independent of $\theta$ and $\alpha$ for
  $i\geq2$.  

  Note that the integral including the boundary terms is
  \begin{equation}
    \begin{cases}
      -\int_0^1 x H_i(x)\,dx=0\\
      -\int_0^1 (1-x) H_i(x)\,dx=0\,;
    \end{cases}
  \end{equation} 
  the boundary terms offset the probability mass in the interior.
\end{rmk}

The first two polynomials are $U_2(x)=-1$ and
$U_3=(2-4x)$; the recurrence relation to calculate all other
polynomials is \citep{Song12}
\begin{equation}\label{eq:rec_rel_small}
   U_{i+1}(x) \frac{(i+1)(i-1)}{2i(2i-1)}
   = U_i(x)\left(x-\tfrac12\right)-U_{i-1}(x)\frac{(i-1)}{2(2i-1)}\,.
 \end{equation}

The $U_i(x)$ solve the differential equation:
\begin{equation}\label{eq:gegen_diff}
  -\lambda_i U_i(x)=\frac{\partial^2}{\partial x^2} U_i(x)\,,
\end{equation}
with 
\begin{equation}
   \lambda_i = i(i-1)\,.
\end{equation}
Thus the $\lambda_i$ are also independent of $\theta$ and $\alpha$ for
$i\geq2$. The $U_i(x)$ are orthogonal with the weight function 
\begin{equation}
   w(x) = x(1-x)\,.
\end{equation}
and the proportionality constant is
\begin{equation}\label{eq:prop_const_gegen}
  \Delta_i=\frac{i-1}{(2i-1)i}\,.
\end{equation}

A probability density defined between zero and one can be represented
by an expansion of the $H_i(x)$:
\begin{equation}\label{eq:series_start}
  f(x)= b_1\delta(x-1) +b_0\delta(x)
  +\sum_{i=2}^{n}\left(c_i H_i(x)\right)\,,
\end{equation}
where 
\begin{equation}\label{eq:boundary_start}
  \begin{cases}
  b_0= \int_0^1 x f(x\given t=0)\,dx\,,\\
  b_1=1-b_0= \int_0^1 (1-x) f(x\given t=0)\,dx\,.
  \end{cases}
\end{equation}
Should $f(x)$ have point masses at the boundaries, these are included
in this integration.  The coefficients $c_i$ can be calculated using
\begin{equation}\label{eq:inner_forw_gegen}
  c_i=\frac1{\Delta_i}\lim_{N\to\infty}\int_{1/N}^{1-1/N} x(1-x)U_{i}(x) f(x)\,dx\,,
\end{equation}
where the limit indicates that the integration includes only the
polymorphic region, i.e., no point masses at the boundaries.

\subsubsection{Pure Drift: Dynamics at the Boundaries}

With the pure drift Moran model, the monomorphic boundaries gain from
the flow out from the polymorphic region. For the boundary at one, the
flow of probability mass out from $(N-1)/N$ to one per unit time,
symbolized by $\frac{\partial }{\partial t} F(1/N)$, is given by the
strength of drift. This is, after the appropriate scaling and taking
the limits,
\begin{equation}\label{eq:bound_1_drift}
  \frac{\partial }{\partial t} F((N-1)/N)=\frac{N-1}{N}\phi((N-1)/N,t)\,,
\end{equation}
and similarly at the other boundary. Since the boundaries are the
only way to lose probability mass from the inside, we also have
\begin{equation}\label{eq:bound_conservation}
    \frac{\partial }{\partial t} (F((N-1)/N)+F(1/N))= -\frac{\partial}{\partial t} 
    \int_{1/N}^{1-1/N} \phi(x,t) \, dx\,,
\end{equation}
where similarly the summation was replaced by the appropriate
integral. Together, we have 
\begin{equation}\label{eq:bound_conservation_N}
    -\frac{\partial}{\partial t} \int_{1/N}^{1-1/N} \phi(x,t)\,dx= \frac{N-1}{N}\phi(1/N,t)+\frac{N-1}{N}\phi((N-1)/N,t)\,.
\end{equation}
Furthermore, it is more likely that proximity to a boundary translates
into preferably exiting through this boundary. A simple set of
boundary conditions that accomplish this is
\begin{equation}\label{eq:bound_conservation_N_0_1}
  \begin{split}
    -\frac{\partial}{\partial t} \int_{1/N}^{1-1/N} \phi(x,t) x\,dx=
    \frac{N-1}{N}\phi((N-1)/N,t)\\
    -\frac{\partial}{\partial t} \int_{1/N}^{1-1/N} \phi(x,t) (1-x)\,dx=
    \frac{N-1}{N}\phi(1/N,t)\,.
  \end{split}
\end{equation}

Equations (\ref{eq:bound_conservation_N_0_1}) imply that the flow out
of the polymorphic region per unit time is equal to the force of drift
at $x=1/N$ and $x=(N-1)/N$, respectively, times the amount present
there, while the probability to exit through a certain boundary is
given by the distance to it.

\subsubsection{Pure Drift: A Different Route to the Solution}

In this subsection, the series expansion used to solve the pure drift
Kolmogorov forward equation is reached via a route, where an expansion
with a general function of time $\tau_i(t)$ is considered, rather than
the usual $e^{-\lambda_i t}$, and where the forward diffusion equation
is integrated using the eigenvectors $U_i(x)$; this strategy also
provides the behavior at the boundaries. This prepares the way for the
solution of the boundary mutation drift model later.

\begin{lem}\label{lem:pure_drift}
  The series expansion 
  \begin{equation}\label{eq:series_alltimes}
    f(x,t)= b_1\delta(x-1) +b_0\delta(x)
    +\sum_{i=2}^{n}\left(\tau_i(t)H_i(x)\right)\,,
  \end{equation}
  where the $\tau_i(t)$ fulfil the dynamic system
  \begin{equation}
    \frac{d}{dt} \tau_i(t)=-\lambda_i\tau_i(t)
  \end{equation} 
  with the starting conditions in equations (\ref{eq:bound_conservation_N_0_1})
  provides the global solution \citep[also incorporating
  boundary terms,][]{Tran14b}, of the
  pure drift forward diffusion equation (\ref{eq:forward_nomut}) in the limit $N\to\infty$.
\end{lem}

\begin{pf}
  The strategy of \citet{Kimu55}, Appendix II, is
  followed.---Integrating the differential equation
  (\ref{eq:gegen_diff}), we get
  \begin{equation}\label{eq:int_bound}
    \begin{split}
      -\lambda_i \int_0^1 U_i(x)\,dx&= \int_0^1  \frac{d^2}{dx^2}x(1-x)
      U_i(x)\, dx\\
      &= \int_0^1  \frac{d}{dx} (x(1-x) \frac{d}{dx} U_i(x)+(1-2x)
      U_i(x)\, dx\\
      &=\left[x(1-x) \frac{d}{dx} U_i(x)+(1-2x)
        U_i(x)\right]_0^1\\
      &=-U_i(0)-U_i(1)\,.
    \end{split}
  \end{equation}
  Conditional on eventual fixation at the boundary one, the forward
  generator is \citep[][section 4.6]{Ewen04}:
  \begin{equation}
    {\cal L}_f^{(1)}=\left(\frac{\partial^2}{\partial
        x^2}x(1-x)\right) -\left(\frac{\partial}{\partial x}(1-x)\right)\,.
  \end{equation}
  Applying this generator to $U_i(x)$ and integrating, results in
  \begin{equation}\label{eq:conditional}
    \begin{split}
      &\int_0^1  \frac{d^2}{dx^2}x(1-x) -\left(\frac{d}{d x}(1-x)\right)
      U_i(x)\, dx\\
      &\qquad= \int_0^1  \frac{d}{dx} \left(x(1-x) \frac{d}{dx} U_i(x)+(1-2x)
      U_i(x) -(1-x) U_i(x)\right)\, dx\\
      &\qquad=\left[x(1-x) \frac{d}{dx} U_i(x)-x
        U_i(x)\right]_0^1=-U_i(1)\,.
    \end{split}
  \end{equation}
  From equations (\ref{eq:int_bound}, \ref{eq:conditional}, and
  \ref{eq:bound_vals}), we obtain for all $i$
   \begin{equation}\label{eq:int_bound_1}
    \begin{split}
      -\lambda_i \int_0^1 xU_i(x)\,dx&= \int_0^1  \frac{d^2}{dx^2}x(1-x) -\left(\frac{d}{d x}(1-x)\right)
      -U_i(x)\, dx\\
      &=-U_i(1)\,.
    \end{split}
  \end{equation}
  Substituting
  \begin{equation}
    f(x,t)= \sum_{i=2}^{n}\left(\tau_i(t)U_i(x)\right)
  \end{equation}
  into
  \begin{equation}
    -\frac{\partial}{\partial t} x\phi(x,t)= -\left(\frac{\partial^2}{\partial x^2}x(1-x)  -\frac{\partial}{\partial x}(1-x)\right)  \phi(t,x)\,,
  \end{equation}
  integrating and taking the limit $N\to\infty$,
  we obtain
  \begin{equation}\label{eq:bound_conservation_infty}
    \begin{split}
    -\frac{d}{d t} \lim_{N\to\infty}\int_{1/N}^{1-1/N} x\phi(x,t)\,dx&=-\frac{d}{dt} \lim_{N\to\infty}\int_{1/N}^{1-1/N}
    \sum_{i=2}^{\infty}\left(\tau_i(t)xU_i(x)\right)\,dx\\
    &=\sum_{i=2}^{\infty}\left(\tau_i(t) U_i(1)\right)\\
    &=\phi(1,t)\,.
    \end{split}
  \end{equation}
  This corresponds to the limit $N\to\infty$ of equation
  (\ref{eq:bound_conservation_N_0_1}) for boundary one.  Combining
  this result with equation (\ref{eq:int_bound_1}), we obtain
  \begin{equation}\label{eq:sum}
    \begin{split}
      -\frac{d}{dt}\sum_{i=2}^{n}\left(\tau_i(t) \frac1{\lambda_i}U_i(1)\right)&=\sum_{i=2}^{\infty}\left(\tau_i(t) U_i(1)\right),.
      \end{split}
  \end{equation}
  The solution of the system of differential equations
  \begin{equation}
    \frac{d}{dt} \tau_i(t)=-\lambda_i\tau_i(t)
  \end{equation}
  fulfils equation (\ref{eq:sum}) for all $i$.  An analogous
  calculation for the boundary at one and summing the results for both
  boundaries, shows that the series expansion using the
  Gegenbauer polynomials fulfils both the pure drift diffusion
  equation as well as the boundary conditions in the limit
  ${N\to\infty}$.  Noting that, with the
  $U_i(x)$ augmented by the boundary terms, whatever leaves the
  polymorphic region for each $H_i(x)$ at $x=1/N$ and $x=(N-1)/N$ in
  the limit ${N\to\infty}$, is added to the monomorphic boundaries at
  $x=0$ and $x=1$, respectively, completes the proof.
\end{pf}

\begin{rmk}
  With the starting conditions, it follows that
  $\tau_i(t)=c_ie^{-\lambda_i t}$, which can also be obtained by
  separation of variables. 
 \end{rmk}

\subsection{No Net-Flow Boundary Condition} 

Substituting the function $eq(x)=x^{-1}(1-x)^{-1}$ into the pure drift
forward equation (\ref{eq:forward_nomut}), shows that $eq(x)$ is a
(local) equilibrium solution:
  \begin{equation}\label{eq:eq(x)_is_local_equil}
    \begin{split}
      \frac{\partial}{\partial t} eq(x)
      &=\frac{\partial^2}{\partial x^2}x(1-x) eq(x)\\
      0&=\frac{\partial^2}{\partial x^2}x(1-x)\,x^{-1}(1-x)^{-1}
      =0\,.
    \end{split}
  \end{equation}
In fact, there is no net flow into or out of an
arbitrary interval $[a,b]$ within $[1/N,(N-1)/]$, as can be deduced
by integration:
\begin{equation}
  \begin{split}
    \int_a^b \frac{\partial}{\partial t} eq(x)\,dx&= \int_a^b  \frac{d^2}{dx^2}x(1-x)
    eq(x)\, dx\\
    &= \int_a^b  \frac{d^2}{dx^2}x(1-x) x^{-1}(1-x)^{-1}\, dx=0.
  \end{split}
\end{equation}
Obviously, $eq(x)$ does not fulfil the boundary conditions in equation
(\ref{eq:bound_conservation_N_0_1}), as the probability mass in the
vicinity of $x=(N-1)/N$ and $x=1/N$ would continually lead to loss by
drift.

Only if this loss is balanced exactly by probability mass continually
replenished from the boundaries, a function proportional to $eq(x)$
may therefore be the polymorphic part of a global equilibrium
solution.  Considering the symmetry of $eq(x)$ and the boundaries
$[1/N,(N-1)/N]$, this process would have to be symmetric.

A population genetic force that may accomplish this is mutation. While
the assumption that in equilibrium mutations from the boundaries
exactly offset the loss through drift at both boundaries may sound
improbable, the next subsection makes just that plausible.

\subsection{The Boundary Mutation-Drift Diffusion Model}

\subsubsection{The Boundary Mutation-Drift Diffusion Model: Slow Time
  Scale; Mutation} 

For the boundary mutation-drift model, we are searching for solutions
for the pure drift Kolmogorov forward equation
(\ref{eq:forward_nomut}) with boundary conditions that include
mutations given some starting density for all times. This model should
approximate the general mutation drift Kolmogorov forward equation
(\ref{eq:Kol_for}) for small scaled mutation rates. For this, a
spectral decomposition is used as before. We make the ansatz
\begin{equation}
  \phi(x,t)=H_0^{\alpha}(x) +\sum_{i=1}^{\infty} \tau_i(t) H_i(x)\,,
\end{equation}
with the eigenvectors $H_i(x)$ identical to those in
(\ref{eq:fast_eigenvectors}) for $i\geq2$.  Continuing with the
strategy of expanding the eigenfunction to zeroth order in $\theta$
and including boundary terms, we obtain for $i=0$
\begin{equation}
   H_0^{\alpha}(x) =w^{\theta,\alpha}(x)R^{\theta,\alpha}(x)=
   \beta\delta(x)+\alpha\delta(x-1)+O(\theta)\,.
\end{equation}
The eigenfunction for $i=1$ can be obtained from equation
(\ref{eq:power_jac_gegen}), such that
\begin{equation}\label{eq:slow_eigenvectors}
  \begin{cases}
    H_0^{(\alpha)}(x)=\beta\delta(x)+\alpha\delta(x-1)\,,\\
    H_1(x)=-\delta(x)+\delta(x-1).
  \end{cases}
\end{equation}
Obviously, these two eigenfunctions are unaffected by the dynamics in
the polymorphic region inside $[1/N,(N-1)/N]$.

Note that the only probability mass of these two eigenfunction is at
the boundaries, such that only eigenvectors with $i\geq2$ have nonzero
probability masses in the polymorphic region. Hence, the model
separates two spatial regions: the monomorphic boundaries and the
polymorphic interior.  The corresponding eigenvectors are
$\lambda_0=0$ and $\lambda_1=\theta$. As $\theta\ll1$ and
the $\lambda_i>1$ for all eigenvalues with $i>2$, two different
temporal regions can be separated, in addition to the two different
spatial regions. Thus, evolution is modeled as a two-time process,
where the slow dynamics of $b_0(t)$ and $b_1(t)$ are evolving
independently from the polymorphic region, while the fast dynamics in
the polymorphic region are in dynamic equilibrium with the slow
dynamics at the boundaries. Generally, we are thus looking at a system
of differential equations, which for the slowly evolving part of the
system is
\begin{equation}\label{eq:dynamics_boundaries}
  \begin{cases}
    \tau_0(t)=1\,,\\
    \frac{d}{dt} \tau_1(t)= - \theta \tau_1(t)\,.
  \end{cases}
\end{equation}
Initially, $b_1(t=0)=\Pr(x=1\given t\to\infty)= \int_0^1 x f(x\given
t=0)\,dx$. The solution over time is
$\tau_1(t)=(b_1(t=0)-\alpha)e^{-\theta t}$, such that the boundary
values will slowly, at a rate of $\theta$, approach the equilibrium
values 
\begin{equation}\label{eq:evol_slow}
  b_1(t)=\alpha +(b_1(t=0)-\alpha)e^{-\theta t}=1-b_0(t)
\end{equation}
Note that $b_0(t)$ and $b_1(t)$ correspond to the probability mass
currently at the boundaries plus the probability mass within the
polymorphic region expected to be fixed by drift at the respective
boundaries. They would only be identical to the probability mass
currently at the boundaries, if there were no probability mass in the
polymorphic region.

\subsubsection{The Boundary Mutation-Drift Diffusion Model: Fast Time
  Scale; Drift and Mutation} 

For small scaled mutation rates, i.e., $\theta\ll 1$, \citet{Vogl12}
suggested to approximate the Moran model presented above by a model,
where the dynamics of polymorphic alleles are only governed by drift,
while mutations only occur in the monomorphic states at the
boundaries, i.e., at $x=0$ or $x=1$. A motivation of this model was
that the probability of a mutation hitting a polymorphic allele is
approximately $2\alpha\beta\theta\log(N)$, which is small if $N$ is
not overly large. Simulations in \citet{Vogl12} show that for the
statistic ``frequency of polymorphism in a sample of size two'' the
approximation holds well for $\alpha\beta\theta<0.01$ (see their
Fig.~1 and note that
$2\theta_0\theta_1/(\theta_0+\theta_0)=\alpha\beta\theta<0.01$). In
the diffusion limit, $N$ is assumed to approach infinity, such that
this argument becomes obsolete and other considerations are needed.

With small scaled mutation rates, the influence of mutations relative
to the effect of drift is small, if $x$ is away from the immediate
vicinity of the boundaries. Mutations affect the mean of $x$
increasing or decreasing it by $1/N$ with probabilities
$\alpha\theta(1-x)$ and $\beta\theta x$, respectively. Compared to the
probability of the same increase or decrease by drift $x(1-x)$, this
is appreciable only close to the boundaries, i.e., close to zero,
where $x$ is equal to or smaller than $\alpha\theta$, or close to one,
where $(1-x)$ is equal to or smaller than $\beta\theta$. In Fig.~1,
the region close to zero is presented for a population in equilibrium
with $\alpha\theta=0.03$ and $\beta\theta=0.015$ (these parameter
values are actually close to the maximum $\alpha\beta\theta$ allowed
by the approximation of small scaled mutation rates). In the Figure,
the rates of the population genetic forces (i.e., mutation and drift)
are multiplied with the equilibrium beta density to show the relative
equilibrium contributions of mutation and drift in different regions
of $x$. It can be seen, that the relative force of drift is almost
constant between zero and one, since the density of $x$ times the
probability of drift is not far from constant, except extremely close
to the boundaries, where it drops sharply to zero. In the vicinity of
zero, the mutational force towards zero has almost no influence (i.e.,
it is indistinguishable from a horizontal line at zero), while the
mutational force towards one is larger than that of drift between zero
and about $\alpha\theta=0.03$ and diminishing from there. For small
$\theta$, the force of drift in equilibrium is approaching a
horizontal line at the level $\alpha\beta\theta$ between zero and one
(excluding the boundaries, where it is zero), while the forces of
mutation approach delta functions at zero and one. 

These considerations are analogous to those in \citet{Step97} and
\citet{Taut00}: a selective force below that of drift has little or no
influence on evolution, analogous to the uncertainty principle in
quantum physics. In our case case, this uncertainty principle is
applied to the force of mutation instead of selection.

As the probability of mutation per Moran event is $\mu$, the scaled
mutation rate per unit of time in the diffusion model becomes
$N^2\mu=N\theta$, such that the mutational terms become
\begin{equation}\label{eq:boundary_terms}
  \begin{cases}
    N\alpha\theta\delta(x)\int_0^1(1-x) \phi(x,t)\,dx &\text{ at 0 and}\\
    N\beta\theta\delta(x-1) \int_0^1x \phi(x,t)\,dx &\text{ at 1.}
  \end{cases}
\end{equation}

While, with the general model, the effects of mutation are incorporated in
the Kolmogorov forward equation (\ref{eq:Kol_for_diff}) by the term
$\theta \frac{\partial}{\partial x} (\alpha-x)$, with the small
scaled mutation model, they are incorporated by the delta functions at
the boundaries (\ref{eq:boundary_terms}):
\begin{equation}\label{eq:forward_mut}
  \frac{\partial}{\partial t} \phi(x,t)
  =\frac{\partial^2}{\partial x^2}x(1-x) \phi(x,t) +N\alpha\theta\delta(1/N-x)b_0(t)+N\beta\theta\delta(x-(N-1)/N) b_1(t)\,,
\end{equation}
with $b_0(t)=\int_0^1 (1-x)\phi(x,t) \,dx$ and $b_1(t)=\int_0^1
x\phi(x,t) \,dx$ as above. This equation implies that the allelic
proportions $x$ are subject to drift everywhere in the polymorphic
region; additionally, mutants arrive at $x=1/N$ and $x=(N-1)/N$ with
rates per generation of $\alpha\theta b_0(t)$ and $\beta\theta
b_1(t)$, respectively.

The boundary conditions analogous to those with pure drift
(\ref{eq:bound_conservation_N_0_1}) are:
\begin{equation}\label{eq:smallmut_bound_0_1}
  \begin{split}
  -\frac{\partial}{\partial t}  \int_{1/N}^{(N-1)/N} x\phi(x,t)\,dx
  &=\frac{N-1}{N}\phi((N-1)/N,t)+N\beta\theta b_1(t)\\
  -\frac{\partial}{\partial t} \int_{1/N}^{(N-1)/N}
  (1-x)\phi(x,t)\,dx
  &=\frac{N-1}{N}\phi(1/N,t)
  +N\alpha\theta b_0(t)\,.\\
  \end{split}
\end{equation}

\subsubsection{The Boundary Mutation-Drift Diffusion Model: General
  Solution}

\begin{thm}
  Starting from a density $f(x)$ within the unit interval
  (eq.~\ref{eq:forward_nomut}) and with the boundary conditions in
  (eq.~\ref{eq:smallmut_bound_0_1}) but letting $N\to\infty$, the
  following function provides the general solution for all times of
  the Kolmogorov forward equation of pure drift diffusion
  \begin{equation}\label{eq:expansion_bound}
    \phi(x,t)=H_0^{(\alpha)}(x)+\sum_{i=1}^\infty \tau_i(t)\,H_i(x)\,,
  \end{equation}
  with the previously defined eigenfunctions
  (eqs. \ref{eq:slow_eigenvectors} and \ref{eq:fast_eigenvectors});
  the $\tau_i(t)$ are given by a system of linear
  inhomogenous first order differential equations
  \begin{equation}\label{eq:dynamics_all}
    \begin{cases}
      \frac{d}{dt} \tau_1(t)= - \theta \tau_1(t)\\
      \frac{d}{dt} \tau_i(t)= 
      -\lambda_i \tau_i(t)-(2i-1)i ((-1)^i\alpha\theta b_0(t)+\beta\theta b_1(t)) 
      \,,\text{ for $i\geq 2$}.\\    
    \end{cases}
  \end{equation}
  The starting values, $\tau_i(t=0)$ for $i\geq 1$, are given by the
  expansion of the initial density $f(x)$ into the eigensystem.
\end{thm}

\begin{pf}
  The slowly evolving part of the system is given in
  (\ref{eq:evol_slow}). For the fast evolving part, note that from
  equation (\ref{eq:inner_forw_gegen}), the coefficients for expanding
  the delta function are:
  \begin{equation}
    \begin{split}
      c_i&=\lim_{N\to\infty}\left(\frac1{\Delta_i}\int_{0}^{1} x(1-x) U_{i}(x)
        N\delta((N-1)/N-x)\,dx)\right)\\
      &=\lim_{N\to\infty}\frac{(N-1)  U_i((N-1)/N)}{N\Delta_i}\\
      &=\frac{U_i(1)}{\Delta_i}\,,
    \end{split}
  \end{equation}
  and analogously for the boundary at zero. Similarly, the incoming
  probability mass needs to be distributed among the eigenfunctions
  proportional to their contributions at the boundaries, which are
  $(-1)^i/i$ at zero $1/i$ at one.
  
  Substituting
  \begin{equation}
    f(x,t)= \sum_{i=2}^{n}\left(\tau_i(t)U_i(x)\right)
  \end{equation}
  into
  \begin{equation}
    -\frac{\partial}{\partial t} x\phi(x,t)=
    -\left(\frac{\partial^2}{\partial x^2}x(1-x)
      -\frac{\partial}{\partial x}(1-x)\right)  \phi(t,x) -N\beta\theta\delta(x-(N-1)/N) b_1(t)\,,
  \end{equation}
  integrating and taking the limit $N\to\infty$,
  we obtain
  \begin{equation}\label{eq:bound_conservation_infty_smallmut_1}
    \begin{split}
    -\frac{d}{d t} \lim_{N\to\infty}\int_{1/N}^{1-1/N} x\phi(x,t)\,dx&=-\frac{d}{dt} \lim_{N\to\infty}\int_{1/N}^{1-1/N}
    \sum_{i=2}^{\infty}\left(\tau_i(t)xU_i(x)\right)\,dx\\
    &=\sum_{i=2}^{\infty}\left(\tau_i(t) U_i(1)\right)+N\beta\theta b_1(t)\frac{U_i(1)}{i\Delta_i}\\
    &=\phi(1,t)+N\beta\theta b_1(t)\,.
    \end{split}
  \end{equation}
  This corresponds to the limit $N\to\infty$ of equation
  (\ref{eq:smallmut_bound_0_1}) for boundary one. 
  This equation leads to 
  \begin{equation}\label{eq:series_smallmut_1}
    \begin{split}
   -\frac{d}{dt} \lim_{N\to\infty}\int_{1/N}^{1-1/N}
    \sum_{i=2}^{\infty}\left(\tau_i(t)\frac1{\lambda_i}U_i(1)\right)\,dx
    =\sum_{i=2}^{\infty}\left(\tau_i(t) U_i(1)\right)+N\beta\theta b_1(t)\frac{U_i(1)}{i\Delta_i}\,.
    \end{split}
  \end{equation}
  The solution of the system of equations
  \begin{equation}\label{eq:third_1}
    \begin{split}
       \frac{d}{dt}\tau_i(t)  =-\lambda_i\tau_i(t) 
    - (2i-1)i  \beta\theta b_1(t)
    \end{split}
  \end{equation}
  fulfils equation (\ref{eq:series_smallmut_1}) for all $i$. 
 
  Analogously, we obtain for the boundary at zero
  \begin{equation}\label{eq:bound_conservation_infty_smallmut_0}
    \begin{split}
    -\frac{d}{d t} \lim_{N\to\infty}\int_{1/N}^{1-1/N} (1-x)\phi(x,t)\,dx&=-\frac{d}{dt} \lim_{N\to\infty}\int_{1/N}^{1-1/N}
    \sum_{i=2}^{\infty}\left(\tau_i(t)(1-x)U_i(x)\right)\,dx\\
    &=\sum_{i=2}^{\infty}\left(\tau_i(t) U_i(0)\right)+N\beta\theta b_1(t)\frac{(-1)^iU_i(0)}{i\Delta_i}\\
    &=\phi(0,t)+N\alpha\theta b_0(t)\,,
    \end{split}
  \end{equation}
  such that eventually
  \begin{equation}\label{eq:third_0}
    \frac{d}{dt}\tau_i(t)  =-\lambda_i\tau_i(t) 
    - (2i-1)i  (-1)^i\alpha\theta b_0(t),.
  \end{equation}
  Summing equations (\ref{eq:third_1}) and (\ref{eq:third_0}), we obtain
  \begin{equation}\label{eq:third_both}
    \frac{d}{dt}\tau_i(t)  =-\lambda_i\tau_i(t) 
    - (2i-1)i  \left((-1)^i\alpha\theta b_0(t)+\beta\theta b_1(t)\right),.
  \end{equation}
  The same considerations as with lemma (\ref{lem:pure_drift})
  complete the proof.
\end{pf}

\begin{rmk}
  Note that the differential equations (\ref{eq:dynamics_all}) for
  $i\geq 2$ can be rearranged to
    \begin{equation}
      \frac{\frac{d}{dt} \tau_i(t)}{\frac{2i-1}{i-1} ((-1)^i\alpha\theta b_0(t)+\beta\theta b_1(t)) 
      + \tau_i(t)}= -\lambda_i\,.\\    
  \end{equation}
  Thus, separation of variables may be used.
\end{rmk}

\subsubsection{The Boundary Mutation-Drift Diffusion Model:
  Equilibrium Solution}

\begin{cor}
  The equilibrium solution of the dynamic system with the slowly
  evolving part given by equation (\ref{eq:evol_slow}) and the
  boundary condition (\ref{eq:smallmut_bound_0_1}) is given by 
  \begin{equation}\label{eq:equil_small_theta}
    \begin{split}
      Eq(x\given \theta,\alpha)=\Pr(x\given\theta,\alpha) &=
      (\beta-\alpha\beta\theta \log(N-1))\delta(x)\\
      &\qquad+\frac{\alpha\beta\theta}{x(1-x)}
      +(\alpha -\alpha\beta\theta \log(N-1))\delta(x-1)\,.
    \end{split}
  \end{equation}
  where the interior region is bounded by $1/N$ and $(N-1)/N$ in the
  limit $N\to\infty$.
\end{cor}

\begin{pf}
  For any starting value, $\tau_1(t\to\infty)=0$, such that
  $b_0(t\to\infty)=\beta$ and $b_1(t\to\infty)=\alpha$. Substituting
  these values into the dynamical system (eq.~\ref{eq:dynamics_all})
  and setting the derivates to zero results in:
  \begin{equation}
      0= 
      -(2i-1)i (\alpha \beta\theta(-1)^i+\alpha\beta\theta) 
      - \lambda_i \tau_i(t)\,.   
  \end{equation}
  From this, it follows that, for all odd $i$, $\tau_i(t=\infty)=0$,
  and, for all even $i$, 
  \begin{equation}\label{eq:equil_taus}
      \tau_i(t\to\infty)= \alpha\beta\theta(4i-2)i/\lambda_i
      = -\alpha\beta\theta(4i-2)/(i-1)\,.\\    
  \end{equation}
  The function
  \begin{equation}
    \phi(x,t\to\infty)=H_0(x)+\alpha\beta\theta\sum_{i=1}^\infty c_{2i}\,H_{2i}(x)
  \end{equation}
  corresponds to the modified Gegenbauer expansion of the equilibrium
  solution for $N\to\infty$ where
  \begin{equation}
    c_{2i}=\frac1{\Delta_{2i}}\int_0^1 x(1-x)U_{2i}(x) x^{-1}(1-x)^{-1}\,dx
    =-\frac{2(4i-1)2i}{2i-1}\frac1{2i}=\frac{4(2i)-2}{2i-1}\,.
  \end{equation}
  Since the function $x^{-1}(1-x)^{-1}$ is symmetric, the boundary
  terms correspond to half the integral over the series expansion,
  which is $\lim_{N\to\infty} 2\alpha\beta\theta \log(N-1))$.
\end{pf}

\begin{rmk}
  $Eq(x\given \theta,\alpha)$ fulfils the boundary conditions in
  (\ref{eq:smallmut_bound_0_1}), also before taking the limit
  $N\to\infty$, as can be shown by substitution. As long as $N$ is not
  too large, $Eq(x\given \theta,\alpha)$ is a proper probability density,
  i.e., everywhere greater than zero and integrating to one over the
  interval. $Eq(x\given \theta,\alpha)$ corresponds to the equilibrium
  solution for the single mutation-drift Moran model \citep{Vogl12}.
\end{rmk}

\subsubsection{Data: Likelihood, Joint and Posterior Densities, and
  the Marginal Distribution with Modified Gegenbauer Polynomials}

The following theorem motivates the interpretation of the boundary
mutation-drift model system using modified Gegenbauer polynomials as a
Taylor series expansion to first order in $\theta$ of the solution of
the general mutation drift model with Jacobi polynomials.

\begin{thm}\label{sample_equil}
  Consider again a sample of size $M$ from a binomial distribution
  (\ref{eq:binomial}) conditional on the allelic proportion $x$, where
  $y$ is the number of alleles of the first type. The probability
  distribution $\Pr(y\given M,\theta,\alpha)$ resulting from a Taylor
  expansion to first order in $\theta$ at $\theta=0$ of the
  beta-binomial compound distribution (eq.~\ref{eq:beta_binomial}),
  where the beta distribution (\ref{eq:equil_beta}) is taken as a
  prior, is identical to the marginal distribution of $y$ resulting
  from taking the equilibrium density $Eq(x\given \theta,\alpha)$
  (equation \ref{eq:equil_small_theta}) as a prior for the allelic
  proportion $x$, and then taking the limit $N\to\infty$ while
  integrating the resulting joint density over $x$.
\end{thm}
\begin{pf}
  The beta-binomial compound distribution (\ref{eq:beta_binomial}) is
  expanded into a power series in $\theta$ at $\theta=0$ up to first
  order. For a polymorphic sample, $1\leq y \leq M-1$, the
  Taylor series expansion of the beta-binomial compound distribution at
  $\theta=0$ is:
  \begin{equation}\label{eq:marg_poly}
    \begin{split}
      \Pr(y\given \theta,\alpha,M)&=
      \binom{M}{y}\,
      \frac{\Gamma(\theta)}{\Gamma(\alpha\theta)\Gamma(\beta\theta)}\,
      \frac{\Gamma(y+\alpha\theta)\Gamma(M-y+\beta\theta)}{\Gamma(M+\theta)}\\
      &=\alpha\beta\theta\, \binom{M}{y}\,
      \frac{\Gamma(y)\Gamma(M-y)}{\Gamma(M)}+O(\theta^2)\\
      &=\alpha\beta\theta\, \frac{M}{y(M-y)}+O(\theta^2)\,.
    \end{split}
  \end{equation} 
  For a monomorphic sample with $y=0$, the derivative of $\Pr(y=0\given
  \theta,\alpha,M)$ with respect to $\theta$ is:
  \begin{equation}
    \begin{split}
      &\frac{d}{d\theta}\Pr(y=0\given \theta,\alpha,M)=\frac{d}{d\theta}
      \left(\frac{\Gamma(\theta)}{\Gamma(\alpha\theta)\Gamma(\beta\theta)}\frac{\Gamma(\alpha\theta)\Gamma(M+\beta\theta)}{\Gamma(M+\theta)}\right)\\
      &\qquad=\frac{d}{d\theta}\left(\frac{\beta\theta(1+\beta\theta)(2+\beta\theta)\cdots(M-1+\beta\theta)}{\theta(1+\theta)(2+\theta)\cdots(M-1+\theta)}\right)\\
      &\qquad=\beta\left(\tfrac\beta{1+\beta\theta}+\tfrac\beta{2+\beta\theta}+\dots+\tfrac\beta{M-1+\beta\theta}-\tfrac1{1+\theta}-\tfrac1{2+\theta}-\dots-\tfrac1{M-1+\theta} \right) \times\\
      &\qquad\qquad\times\frac{(1+\beta\theta)(2+\beta\theta)\cdots(M-1+\beta\theta)}{(1+\theta)(2+\theta)\cdots(M-1+\theta)}\,.
    \end{split}
  \end{equation}
  Thus the Taylor series expansion at $\theta=0$ is to first order:
  \begin{equation}\label{eq:marg_mono}
    \Pr(y=0\given
    \theta,\alpha,M)=\beta-\alpha\beta\theta\,\sum_{y=1}^{M-1} \frac1y+O(\theta^2)\,,
  \end{equation}
  and analogously for $\Pr(y=M\given \theta,\alpha,M)$.
  
  For polymorphic samples, the joint density of the binomial and
  the equilibrum density $Eq(x\given \theta,\alpha)$ (equation
  \ref{eq:equil_small_theta}) is
  \begin{equation}\label{eq:joint_poly}
    \Pr(1\leq y\leq M-1,x\given\alpha,\theta,M)= \alpha\beta\theta
    \binom{M}{y} x^{y-1}(1-x)^{M-y-1}\,.
  \end{equation}
  Integrating over $x$ and taking the limit $N\to\infty$, such that
  the integration boundaries are 0 and 1, respectively, results in the
  marginal distribution:
  \begin{equation}\label{eq:joint_poly_2}
    \begin{split}
      \Pr(1\leq y\leq M-1\given\alpha,\theta,M)&=\int_0^1
      \alpha\beta\theta x^{y-1}(1-x)^{M-y-1}\,dx\\
      &= \alpha\beta\theta\, \binom{M}{y}\,
      \frac{\Gamma(y)\Gamma(M-y)}{\Gamma(M)}\\
      &=\alpha\beta\theta\, \frac{M}{y(M-y)}\,.
    \end{split}
  \end{equation}
  This is identical to the first order expansion (\ref{eq:marg_poly}).

  For a monomorphic sample, e.g., $y=0$, the joint density is
  \begin{equation}\label{eq:joint_mono}
    \Pr(y=0,x\given\alpha,\beta)=\beta+\alpha\beta\theta (-\log(N-1)+(1-x)^{-1}x^{M-1})\,.
  \end{equation}
  Using the series expansion $(1-x)^{-1}=\sum_{i=0}^\infty x^i$
  results in
  \begin{equation}
    \begin{split}
      -\log(N-1)+\int_{1/N}^{1-1/N}(1-x)^{-1}x^{M}\,dx&=\int_{1/N}^{1-1/N}
      -(1-x)^{-1}+(1-x)^{-1}x^{M-1}\,dx\\
      &=\int_{1/N}^{1-1/N} \sum_{i=0}^{\infty}(-x^i+x^ix^{M-1})\,dx\\
      &= -\int_{1/N}^{1-1/N} \sum_{i=1}^{M-1}x^{i-1}\,dx\\
      &=-\sum_{i=1}^{M-1}\frac{(1-1/N)^i-(1/N)^i}i\,.
    \end{split}
  \end{equation}
  In the limit $N\to\infty$, this converges to
  $-\sum_{i=1}^{M-1}1/i$. The marginal distribution of the monomorphic
  sample then is
  \begin{equation}
    \Pr(y=0\given\alpha,\beta)=\beta- \alpha\beta\theta \sum_{i=1}^{M-1}1/i\,.
  \end{equation}
  This is identical to the first order expansion
  (\ref{eq:marg_mono}). The analogous calculation for $y=M$ completes
  the proof.
\end{pf}

\begin{rmk}
  For polymorphic samples, the joint density
  (\ref{eq:joint_poly_2}) is a polynomial that can be represented without
  loss by the modified Gegenbauer polynomials, as long as $M\leq N$. As long as
  \begin{equation}
    max(\alpha,\beta)\cdot\theta\,\sum_{y=1}^{M-1} \frac1y\leq1\,,
  \end{equation}
  the following joint probability for monomorphic samples, for $y=0$:
  \begin{equation}\label{eq:joint_mono_2}
    \Pr(y=0\given\alpha,\beta)=\beta
    +\alpha\beta\theta \left(-\sum_{i=1}^N\frac1i+\sum_{i=M}^N x^{i-1}\right)
  \end{equation}
  and analogously for $y=M$, results a proper joint density.
  This is also a polynomial in $x$ and can therefore be represented
  without loss using the modified Gegenbauer polynomials. Note that
  polynomials can generally be represented as a linear combination of
  beta densities with integer parameters. Furthermore, the order
  of the expansion $N$ in effect takes the role of the effective
  population size, with the series expansion.
\end{rmk}

\subsubsection{Numerics}

With the statistical language ``R'' (``www.r-project.org'') and its
high-precision algebra package ``Rmpfr'', the terms of the modified
Gegenbauer polynomials up to the order $50$ can be calculated within
minutes using this method. With an expansion of order $N$, the
beta-binomial posterior distributions (equation \ref{eq:post}) of
samples of size $N$ can be represented exactly.

With terms up to $i=50$, the equilibrium expansion is shown in Fig.~2.
Note that an expansion using $d_i$ as coefficients results in an
approximation proportional to the delta function at zero or one (also
shown in Fig.~2). Further approximations to beta densities that
arise in the analysis of real data are also shown in Fig.~2.

\subsubsection{Example: A Change in the Mutation Bias with Modified
  Gegenbauer Polynomials}

For short introns, \citet{Clem12a} argue that in {\em Drosophila
  melanogaster} a change in mutation bias from mildly to strongly
biased towards AT over GC can explain the observed pattern of site
frequency spectra. The model they used for analyses was based on
quasi-equilibrium depending on the frequencies at the boundaries. In
this subsection, a more precise model is investigated.

Suppose that the mutation bias changes from $\alpha_a(t=0)=1/3$ to
$\alpha_c(t>0)=2/3$, while $\theta$ remains constant. We want to
obtain the prior density analogous to the equilibrium density
$(x(1-x))^{-1}$ at an arbitrary time $t$ therafter. Initially,
$b_0(t=0)=2/3$, $b_1(t=0)=1/3$, while the function in the interior is
$2/9\,\theta\,x^{-1}(1-x)^{-1}$. At time $t=0$, the equilibrium
starting condition can be expanded to:
\begin{equation}\label{eq:initial_equil_expansion}
  H_0^{\alpha_a}(x)+\sum_{i=2}^{\infty}\bar c_i H_i(x)\approx
  \tfrac{2}{3}\,\delta(0)+\tfrac{1}{3}\,\delta(1)+\tfrac{2}{9}\,\theta\,\sum_{i=2}^{\infty} \frac{( (-1)^i+1) (2i-1)}{i-1} H_i(x)\,.
\end{equation}
Set the $\tau_i(0)=\bar c_i$. Considering first the slow dynamics,
which are independent from the fast dynamics, $b_1(t)$ will eventually
increase from $\tfrac13$ to $\tfrac23$:
\begin{equation}
  b_1(t)= \frac{2-\,e^{-\theta t}}3\,.
\end{equation}

The coefficients of the interior eigenfunctions evolve according to a
linear inhomogenous first order differential equation
(eq. \ref{eq:dynamics_all}):
\begin{equation}\label{eq:inhom}
  \begin{split}
  \tfrac{d}{dt} \tau_i(t)&=-\tfrac23\theta (\tfrac13+\tfrac13\,e^{-\theta t}
  )(-1)^i(2i-1)i-\tfrac13\theta (\tfrac23-\tfrac13\,e^{-\theta t}) (2i-1)i-\lambda_i
  \tau_i(t)\\
  &=-\tfrac29\theta ((-1)^i+1)(2i-1)i -\tfrac19(2(-1)^i-1)(2i-1)i\,e^{-\theta t}-\lambda_i
  \tau_i(t)\,.
  \end{split}
\end{equation}
With the starting conditions $\tau_i(0)=\bar c_i$, the solution to the
differential equation (\ref{eq:inhom}) is, for odd $i$
\begin{equation}
  \tau_i(t)= \tfrac39(2i-1)i\frac{e^{-\theta t}-e^{-\lambda_i t}}{\lambda_i-\theta}\,.
\end{equation}
and for even $i$
\begin{equation}
  \tau_i(t)= -\frac{4(2i-1)i}{9\lambda_i}-\tfrac19(2i-1)i\frac{e^{-\theta t}-e^{-\lambda_i t}}{\lambda_i-\theta}\,.
\end{equation}
Note that even though $\theta$ does not increase and the equilibrium
density is identical before and after the change in mutation bias, the
even eigenfunctions and thus also the probability mass in the interior
increase transiently. Since $\theta\ll \lambda_i$, especially for
higher $i$, a quasi-equilibrium will result rapidly. A graph of the
time course of the modified Gegenbauer expansion of the equilibrium
density with $\theta=0.01$ is presented in Fig.~3.

\section{Summary and Conclusion}

In this article, the starting point is the general biallelic
mutation-drift diffusion equation with two parameters, the scaled
mutation rate $\theta=\mu N$, where $\mu$ is the mutation rate per
reproduction event and $N$ the haploid effective population size, and
the allelic mutation bias $\alpha=\mu_1/\mu=1-\beta$. The
evolution of the population allelic proportion $x$ over the
appropriately scaled time can be found by expanding into a series of
modified Jacobi polynomials \citep[e.g.,][]{Grif10,Song12}. The
equilibrium density corresponds to a beta \citep{Wrig31}. If the
parameters change, e.g., if the mutation bias changes or the
population shrinks or grows, the Jacobi expansion needs to be
changed. For continually changing parameters, this is cumbersome.

With small scaled mutation rates $\theta\ll1$, the interior dynamics
are governed by drift and are relatively fast, while mutations
influence the dynamics mainly at the boundaries at a relatively slow
rate. This fact was already used in much of population genetics theory
(e.g., for deriving the Ewens-Watterson estimater of $\theta$).
\citet{Gute09} used the same approximation in their program $\delta a
\delta i$. In analogy to the discrete model \citep{Vogl12}, a model
with mutations only from the boundaries is developed. The equilibrium
solution of this boundary-mutation drift model has an interior part of
$\alpha\beta\theta\,x^{-1}(1-x)^{-1}$ \citep[see also][]{RoyC10},
while the allelic proportions at the boundaries are influenced only by
the mutation bias. For small $\theta$, the beta-binomial compound
distribution, which results from the general model, can be expanded to
first order in $\theta$ to result in a marginal distribution. The same
marginal distribution is obtained with the use of the
boundary-mutation drift equilibrium density, after taking the limit
$N\to\infty$. For the temporal part, a system of linear differential
equations is derived that corresponds to the general solution of the
boundary-mutation drift model. This solution using the orthogonal
Gegenbauer polynomials seems to correspond to the numeric solution
using a grid in $\delta a \delta i$ \citep{Gute09}, who presumably
assumed unbiased mutations. Since in equillibrium the joint
density of the allelic proportion $x$ given a sample of moderate size
$M$ is proportional to a beta density with integer coefficients, a
polynomial of order $M-1$ for a polymorphic sample, the solution
presented here also has the advantage of producing the exact joint and
posterior densities. Furthermore, the use of orthogonal polynomials connects
to other, earlier theoretical work. In contrast to using Jacobi
polynomial expansions, which are applicable also to large scaled
mutation rates with $\alpha\beta\theta>0.01$, the Gegenbauer
polynomial expansion does not require a change of the basis if
parameters change, e.g., because populations grow or shrink, and is
thus more convenient, when the assumption of small scaled mutation
rates can be justified.

Additionally considering directional selection, as \citet{Gute09} and
\citet{Vogl12} have done for their models, is an obvious
generalization of the approach in this article. \citet{Song12} provide
as a starting point their model and analysis with general mutation
rates.

\section{Acknowledgements}

The author expresses his thanks to the participants in the doctorate
college ``population genetics'' funded by the FWF for stimulating this
research, to Lee Altenberg for an inspiring discussion, and to Andreas
Futschik and two anonymous reviewers for critically reading an earlier
version of manuscript.  \newpage

\bibliography{/Users/vogl/TeX/coal/coal} 

\bibliographystyle{natbibgen}

\newpage 
\paragraph{Figure~1} Comparison of the influence of drift (thick line)
and of a mutation towards allele one (greater than 0) or towards
allele 0 (horizontal line at $y=0$).

\begin{figure}[ht]
\begin{center}
\includegraphics[scale=1]{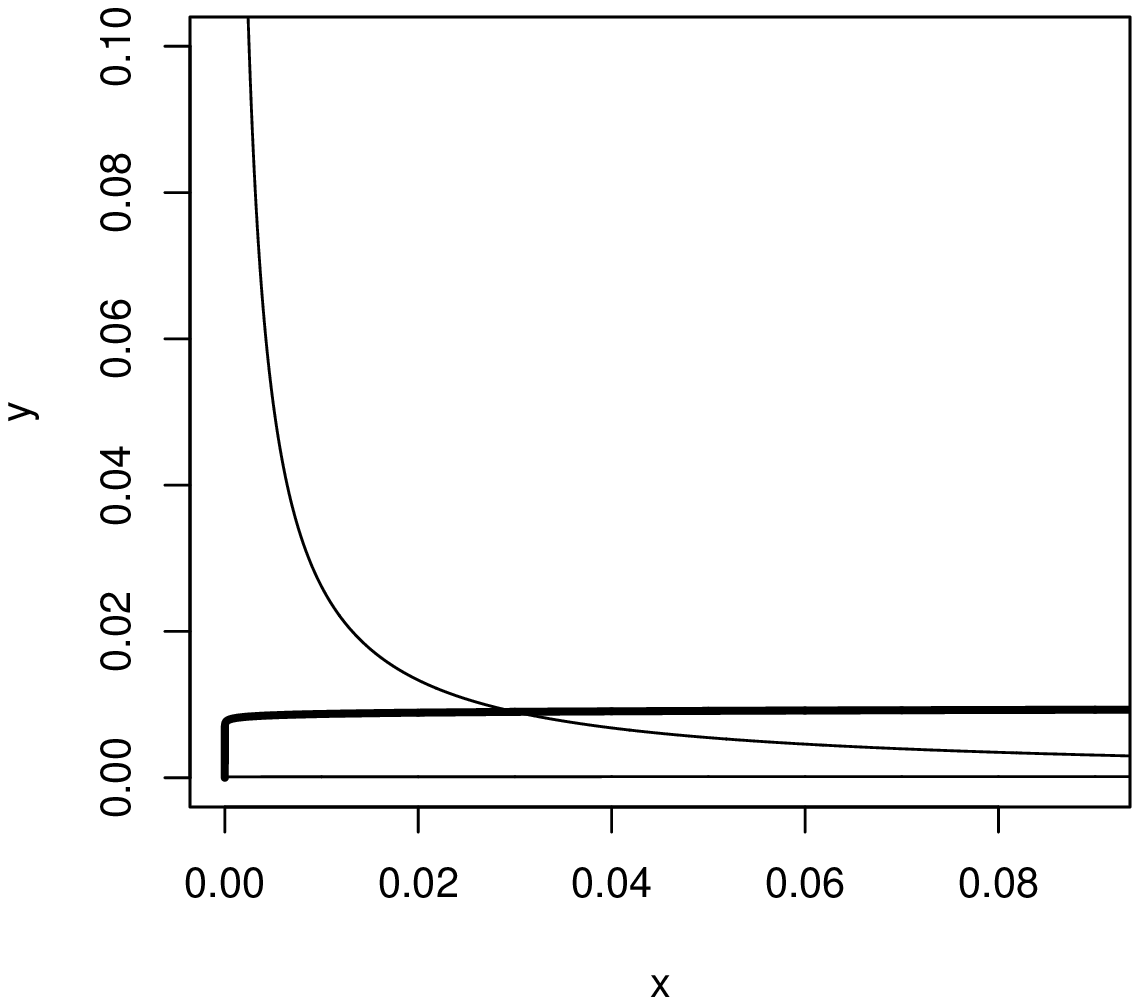}
\end{center}
\end{figure}

\newpage 
\paragraph{Figure~2} Approximate densities using the Gegenbauer
polynomial expansion with terms up to $i=52$. A) Approximation
proportional to the sum of the Dirac delta function
$\lim_{N\to\infty}\left(\frac1{\Delta_i}\int_{0}^{1} x(1-x)
  U_{i}(x)N\delta((N-1)/N-x)\,dx)\right))$ at one and that at zero; B)
approximation to the equilibrium improper density $x^{-1}(1-x)^{-1}$
(wiggliy line) and the function $x^{-1}(1-x)^{-1}$ (smooth line); C)
approximation to the joint posterior density for a sample with
$y=1$, $M=1$ (wiggly line) and the joint density
$2\,x^{1-1}(1-x)^{1-1}$ (smooth line); D) approximation to the joint
posterior density for a sample with $y=3$, $M=6$ (wiggly line)
and the joint density $\binom{6}{3}\,x^{3-1}(1-x)^{3-1}$ (smooth
line).

\begin{figure}[ht]\begin{center}
\includegraphics[scale=0.5]{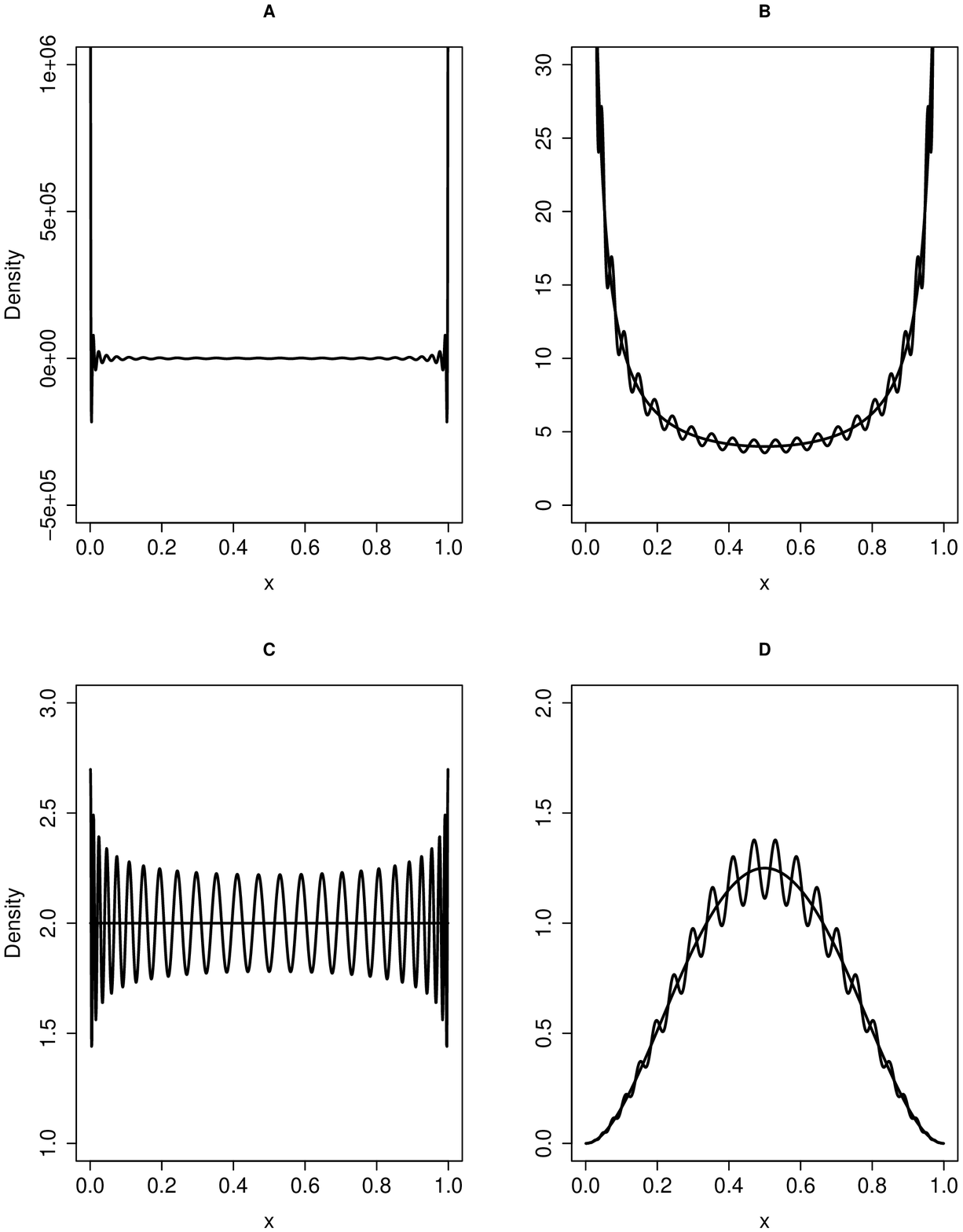}
\end{center}
\end{figure}

\newpage 
\paragraph{Figure~3} The time course of the polymorphic part of the
allele proportions $x$ after a change in the mutation bias. The thin
line represents the improper equilibrium distribution
$x^{-1}(1-x)^{-1}$. The time is (A) $t=0$, (B) $t=0.1$, (C) $t=1$, and
(D) $t=100$.

\begin{figure}[ht]\begin{center}
\includegraphics[scale=0.5]{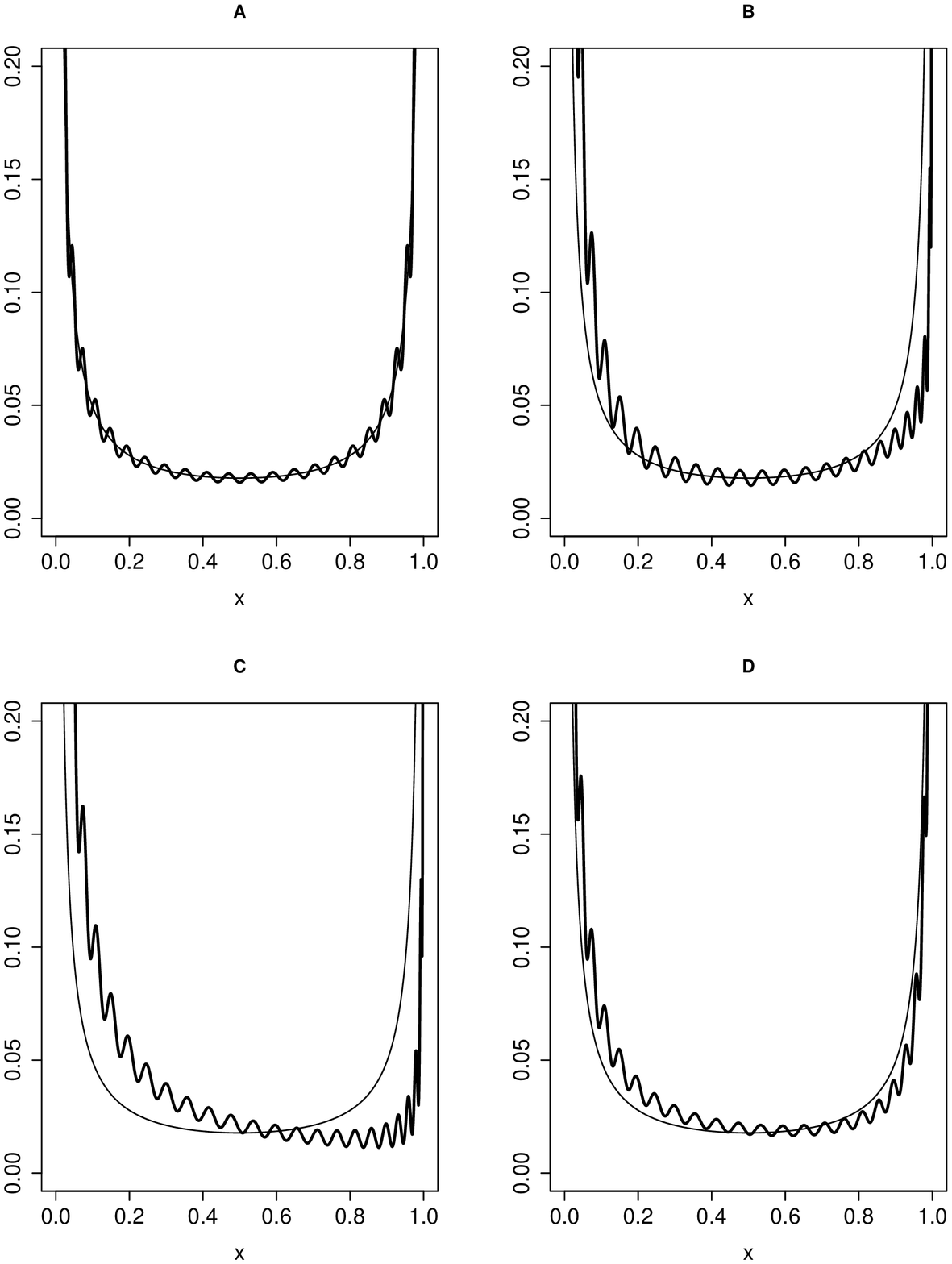}
\end{center}
\end{figure}

\end{document}